\documentclass[11pt,a4paper]{article}
\usepackage{jheppub}
\usepackage{amssymb,amsmath,amsfonts,mathrsfs}
\usepackage{graphicx}
\usepackage[dvipsnames]{xcolor}
\usepackage{verbatim}
\usepackage{epsfig}
\usepackage{color}
\usepackage{multirow}
\usepackage{comment}
\usepackage{datetime}
\usepackage{slashed}
\usepackage{ulem}
\usepackage{framed}
\usepackage{longtable}
\usepackage[mathscr]{euscript}
\usepackage{cancel}
\usepackage{tensor}
 \usepackage{mathtools}
  \usepackage{caption}
  \usepackage{subcaption}
 \usepackage[multiple]{footmisc}
 \usepackage{pgfplots}
 \usepackage{natbib}
\usepackage[applemac]{inputenc}
\usepackage{tikz}
\usepackage{placeins}
\usetikzlibrary{decorations.markings}

\newcommand{\nn}{\nonumber}

\def\be{\begin{equation}}
\def\ee{\end{equation}}
\def\bea{\begin{align}}
\def\eea{\end{align}}

\def\a{\alpha}				\def\g{\gamma}		\def\d{\delta}
	\def\z{\zeta}					\def\q{\theta}
		\def\k{\kappa}		\def\l{\lambda}		
\def\n{\nu}			\def\x{\xi}						\def\r{\rho}	
								
\def\c{\chi}						

		\def\D{\Delta}

\title{Charge instability of JMaRT geometries}
\author{Massimo Bianchi,}
\author{Carlo Di Benedetto,}
\author{Giorgio Di Russo,}
\author {Giuseppe Sudano}

\affiliation{Dipartimento di Fisica,  Universit\`a di Roma ``Tor Vergata"  \& Sezione INFN Roma2, Via della ricerca scientifica 1, 00133, Roma, Italy}

\abstract{We perform a detailed study of linear perturbations of the JMaRT family of non-BPS smooth horizonless solutions of type IIB supergravity beyond the near-decoupling limit. In addition to the unstable quasi normal modes (QNMs) responsible for the ergo- region instability, already studied in the literature, we find a new class of `charged' unstable modes with positive imaginary part, that can be interpreted in terms of the emission of charged (scalar) quanta with non zero KK momentum. We use both matched asymptotic expansions and numerical integration methods. Moreover, we exploit the recently discovered correspondence between JMaRT perturbation theory, governed by a Reduced Confluent Heun Equation, and the quantum Seiberg-Witten (SW) curve of $\mathcal{N} = 2$ SYM theory with gauge group SU(2) and $N_f = (0,2)$ flavours.}

\begin{document}
\maketitle
\flushbottom 
\section{Introduction}
JMaRT geometries represent the first known family of smooth horizonless but non supersymmetric solutions of the 3-charge system \cite{Jejjala:2005yu}. The solutions depend on two\footnote{Actually three, if one acts with an orbifold of order $k$.} integer parameters $m$, $n$, three boost parameters $\delta_i$ and a mass scale $M$. These determine three charges $Q_i$, two angular momenta $J_\phi, J_\psi$ and the ADM mass.
Due to their over-rotation, JMaRT solutions cannot be considered {\it bona fide} micro-states of 3-charge `large' black-holes (BHs) in $D=5$. Moreover due to the presence of an ergo-region and the absence of a horizon, they have been argued and then shown to be unstable \cite{Cardoso:2005gj} under small (linearized) scalar perturbations. Although this instability is well established for self-gravitating objects, such as stars, with an ergoregion and no horizon \cite{Chandrasekhar:1984siy}, the precise physical origin of the instability of JMaRT is not completely clear. The method used to display the instability is to identify Quasi Normal Modes (QNM) with positive imaginary part of the frequency.
Several possibilities were listed in the conclusions of \cite{Jejjala:2005yu}.
Notwithstanding the absence of a horizon, \cite{Chowdhury:2007jx} argued for an interpretation in the form of some kind of Hawking radiation. On the other hand, thanks to the very presence of an ergoregion, Penrose process was also argued to take place in JMaRT geometries \cite{Bianchi:2019lmi}. The end-point of the instability is expected to be a supersymmetric solution of the GMS family \cite{Giusto:2004id, Giusto:2004ip} that corresponds to the choice $m=n+1$ (or equivalent ones). In turn the latter have been shown to be stable under linear scalar perturbations in \cite{Chakrabarty:2019ujg} but argued to suffer some non-linear instability \cite{Eperon:2016cdd}. No direct proof or explicit description of the `decay' process have been given so far. Moreover most of the analysis of the instability has been confined to the near-decoupling limit $Q_p \ll Q_1,Q_5$ and to `neutral' objects with vanishing Kaluza-Klein momentum $p=\lambda/R=0$ (with $\lambda$ an integer).

The aim of the present paper is three-fold.

First we will study what may be termed `JMaRT charge instability' {\it i.e.} `spontaneous emission' of charged (scalar) quanta with non zero KK momentum $p\neq 0$. Focusing on perturbations with $\ell=0$ we avoid confusion with the `ergoregion instability' considered in \cite{Cardoso:2005gj} and find that KK-charged QNMs exist with positive imaginary part. These are the counterpart of the charged particle emission by the JMaRT microstate, that being non BPS ($M_{ADM} > |Q_p|+|Q_1|+|Q_5|$) can reduce its mass and KK charge in the form of (non-)BPS states with $m_p\ge|p|$. Assuming the validity of the `Weak Gravity Conjecture' (WGC), states with $m_p\le |p|$ should exist too that should allow a complete discharge of the solution \cite{Shiu:2016weq}. 

Second, we will not work in the near-decoupling limit, {\it i.e.} we will not assume $Q_p \ll Q_1, Q_5$: in fact we will often consider  $Q_p=Q_1=Q_5 = Q$  for simplicity.  

Third, we will employ the new approach to BH and fuzzball perturbation theory based on the correspondence with quantum Seiberg-Witten (SW) curves of ${\cal N}=2$ SYM theory with gauge group $SU(2)$ and $N_f=(N_L,N_R)$ flavours of fundamental hypers \cite{Aminov:2020yma, Consoli:2022eey, Bianchi:2021mft, Bianchi:2021xpr, Bonelli:2021uvf, Bonelli:2022ten}. JMaRT perturbations are governed by the Reduced Confluent Heun Equation (RCHE) with two regular and one irregular singularity, that corresponds to $N_f=(2,0)$ for the radial part as well as for the (polar) angular part. We will check our results with those available in \cite{Cardoso:2005gj}. Note that in \cite{Cardoso:2007ws} a different branch of BH solutions with $M\ge(a_1+a_2)^2$ was studied and shown not to suffer charge super-radiance.

The paper is organized as follows. 
In Section \ref{sect2} we briefly describe the general solution and its properties. 
In Section \ref{sect3} we describe the separation of the dynamics and study critical geodesics.
In Section \ref{sect4} we study the charged scalar wave equation, its separation and the identification of the relevant parameters.
In section \ref{sect5} we briefly recap the connection between quantum SW curves and BHs/fuzzballs perturbation theory. We identify dictionaries and compute the cycles both for radial and angular wave equation. 
In section \ref{sect6} we then compute the QNMs and show that even for $\ell=0$ but $p\neq 0$ there are QNMs with positive imaginary part. We give our interpretation of the charge instability and discuss possible `phenomenological' implications. 
Section \ref{sect7} contains our conclusions and outlook also in relation with other smooth horizonless geometries such as `topological stars' or `Schwarzschild topological solitons' \cite{Bah:2020pdz, Heidmann:2022ehn}. 
We also (re)compute in appendix \ref{appA} QNMs with $\ell \neq 0$ for comparison with the literature.

\section{JMaRT solution}\label{sect2}

In order to fix the notation and to be self-contained, we briefly review JMaRT solutions and their basic properties. 
JMaRT solutions \cite{Jejjala:2005yu} are (non)-BPS smooth horizonless geometries sourced by three charges: (smeared) D1-branes wrapping a compact $S^1_y$ circle, Kaluza-Klein (KK) momentum along this direction and D5-branes wrapping $S^1_y \times T^4$. The metric depends on the three charges 
\begin{equation} \label{eq:charges}
Q_i = M c_i s_i, \qquad i = 1, 5, p
\end{equation}
where $M$ is a mass parameter and $c_i=\cosh\delta_i$, $s_i=\sinh\delta_i$, $\delta_i \geq 0$. The ADM mass and the angular momenta of this family of solutions read
\begin{equation}
 M_{\mathrm{ADM}} = \frac{M}{2} \bigl (\cosh{2 \delta_1} + \cosh{2 \delta_5} + \cosh{2 \delta_p} \bigr)
\end{equation}
\begin{equation}
J_\psi = - M \bigl (a_1 \cosh \delta_1 \cosh \delta_5 \cosh \delta_p - a_2 \sinh \delta_1 \sinh \delta_5 \sinh \delta_p      \bigr ) 
\end{equation}
\begin{equation}
J_{\phi} = - M \bigl (a_2 \cosh \delta_1 \cosh \delta_5 \cosh \delta_p - a_1 \sinh \delta_1 \sinh \delta_5 \sinh \delta_p      \bigr )  
\end{equation}
where $a_1$ and $a_2$ are two length parameters. Let us notice that exchanging $a_1$ and $a_2$ is tantamount to exchanging the two angular momenta $J_{\phi}$ and $J_{\psi}$. Without loss of generality, we can assume $a_1 \geq a_2 \geq 0$. 

The 6-dimensional metric\footnote{We neglect the dynamics on  $T^4$ that decouples completely for $Q_1=Q_5$.} is
\begin{align}\label{eq:metric}
ds^2=&{1\over \sqrt{H_1 H_5}}\Big\{-(f-M)[d\tilde{t}-(f-M)^{-1}M c_1 c_5(a_1 \cos^2\q d\psi+a_2 \sin^2\q d\phi)]^2\\\nn
&+f[d\tilde{y}+f^{-1}M s_1 s_5(a_2 \cos^2\q d\psi+a_1\sin^2\q d\phi)]^2\Big\}\\\nn
&+\sqrt{H_1 H_5}\Big\{{r^2dr^2\over(r^2+a_1^2)(r^2+a_2^2)-M r^2}+d\q^2\\\nn
&+(f(f-M))^{-1}[(f(f-M)+f a_2^2\sin^2\q-(f-M)a_1^2\sin^2\q)\sin^2\q d\phi^2\\\nn
&+2Ma_1 a_2\sin^2\q \cos^2\q d\psi d\phi\\\nn
&+(f(f-M)+fa_1^2\cos^2\q-(f-M)a_2^2\cos^2\q)\cos^2\q d\psi^2]\Big\}
\end{align}
where $\tilde{t}=t c_p-y s_p$, $\tilde{y}=y c_p-t s_p$ are the `boosted' time and circle coordinates and
\be
f=r^2+a_1^2\sin^2\q+a_2^2\cos^2\q\quad,\quad H_i=f+Ms_i^2
\ee
In terms of $r^2$ the two roots of the denominator of the $g_{rr}$ component of the metric are
\begin{equation}\label{r2+-}
r_\pm^2 = \frac{M - a_1^2 - a_2^2 \pm \sqrt{\bigl (M - a_1^2 - a_2^2 \bigr)^2 - 4 a_1^2 a_2^2}}{2}
\end{equation}
Requiring the absence of curvature singularities, horizons and closed-time-like curves amounts to imposing the following constraints on the parameters of the solution \cite{Bianchi:2019lmi,Bianchi:2021mft,Bianchi:2020yzr}:
\begin{align}
M=a_1^2+a_2^2-a_1 a_2 {c_1^2c_5^2c_p^2+s_1^2s_5^2s_p^2\over  c_1 s_1 s_5 c_5 s_p c_p}\quad, \quad R_y= {M s_1 c_1 s_5 c_5(s_1 c_1 s_5 c_5 s_p c_p)^{1/2}\over \sqrt{a_1 a_2}(c_1^2c_5^2 c_p^2-s_1^2s_5^2s_p^2)}
\label{mR}
\end{align}
so that 
\begin{equation}
r_-^2 < r_+^2 = - a_1 a_2 \frac{s_1s_5s_p}{c_1c_5c_p} < 0
\end{equation}
Moreover, two quantization conditions should hold:
\be
{{\lambda}+{\lambda}^{-1}\over {\sigma}+{\sigma}^{-1}}=m-n,\quad {{\lambda}-{\lambda}^{-1}\over {\sigma}-{\sigma}^{-1}}=m+n
\label{js}
\ee
where\footnote{In order to avoid confusion with the various $s_i$ we denoted by ${\sigma}$ and ${\lambda}$ the parameters $s$ and $j$ of \cite{Jejjala:2005yu}.} we have defined the dimensioless parameters
\be
{\lambda}=\sqrt{a_2\over a_1}\leq1\quad,\quad {\sigma}=\sqrt{s_1 s_5 s_p\over c_1 c_5 c_p}\leq1
\label{parameters}
\ee
\\Setting 
\be
\beta=(m-n+1)(m-n-1)(m+n+1)(m+n-1)
\ee
and assuming ${\lambda}\leq1$ and ${\sigma} \leq1$ the solutions of the system $(\ref{js})$ are given by
\begin{align}
&{\lambda}^2=\frac{\left(m^2-n^2+1-\sqrt{\beta}\right)^2(m^2+n^2-1-\sqrt{\beta})}{8nm^3}\\&{\sigma}^2=\frac{\left(m^2+n^2-1-\sqrt{\beta}\right)}{2nm}
\end{align}

It is useful to write some parameters in terms of $M, m, n$. Inverting $(\ref{mR})$ wrt $a_1, a_2$ and then replacing in $(\ref{r2+-})$ one finds
\begin{align}
&a_{1}=\sqrt{M}\frac{4m^{2}n }{[(m-n)^{2}-1][(m^2-n^2+1-\sqrt{\beta})((m+n)^2-1-\sqrt{\beta})]}\\
&a_{2}=\sqrt{M}\frac{m (m^2-n^2-1-\sqrt{\beta})}{\sqrt{(m-n)^{2}-1}\left[(m+n)^2-1-\sqrt{\beta}\right]}\\
&r_+^2=-a_1a_2\sigma^2=\frac{M}{2} \frac{(1-m^2-n^2+\sqrt{\beta})}{\beta}\\ &r_{-}^{2}=-\frac{a_1a_2}{\sigma^2}=-4M \frac{m^2n^2}{\left[(m-n)^2-1\right]\left[1-(m+n)^2+\sqrt{\beta}\right]^2}\\
&R_y=\sqrt{2M} {mn\sqrt{\beta}\sqrt{m^2+n^2-1-\sqrt{\beta}}\over m^2 \left[2(n^2+1)-m^2+\sqrt{\beta}\right]-(n^2-1)(n^2-1-\sqrt{\beta})}\left(s_1 s_5 \over c_p\right)
\end{align}

The other `observables' $M_{ADM}, J_a, Q_i$ depend on combinations of the `boost' parameters that cannot be expressed only in terms of $M, m, n$.

\section{Geodesic motion}\label{sect3}
In order to gain further insights into the origin of the `charge' instability, we now consider geodesics for massless particles in the JMaRT geometries that obviously satisfy $ds_{(6)}^2=0$.

The large amount of symmetry allows to separate the dynamics, then it is convenient to work in the Hamiltonian formalism with
\be
\mathcal{H}={1\over 2}g^{\mu\nu}P_\mu P_\nu=0
\ee
where $P_\mu = g_{\mu\nu} \dot{x}^\nu$. 

Exploiting the conservation of the momenta $P_t=-E$, $P_\phi=J_\phi$, $P_\psi=J_\psi$ and $P_y$ and denoting by $K^2$ a (positive for $E^2\ge P_y^2$) separation constant\footnote{In fact $K^2\ge (E^2-P_y^2) {\rm min}\{a_1^2, a_2^2\} = (E^2-P_y^2)a_2^2$.}  the polar angular equation reads:
\begin{equation}\label{eqgeodA}
P_\q^2+{J_\phi^2 \over \sin\q^2}+{J_\psi^2 \over \cos\q^2}+(a_2^2 \sin^2\q + a_1^2 \cos^2\q)(E^2-P_y^2)=K^2
\end{equation}
Very much as Carter's constant for Kerr BHs \cite{Chandrasekhar:1984siy}, $K^2$ plays the role of total angular momentum including frame-dragging. 

We can introduce the `physical' radial coordinate 
\be 
\rho=\sqrt{r^2-r_+^2}
\ee
such that the cap is mapped in $\r=0$. In this coordinate, the radial equation reads
\be\label{qgeojmart}
P_\r^2 =Q_\r={ (E^2-P_y^2){\cal R}_3(\r^2)\over\r^2(\r^2+r_+^2-r_-^2)^2}
\ee
where
\be
\label{geopoly}
{\cal R}_3(\r^2)=\r^6+B \r^4+C \r^2+D
\ee
is a cubic polynomial in $\r^2$. Its coefficients can be expressed in terms of the three impact parameters 
\be
b_{\phi,\psi}={J_{\phi,\psi}\over \sqrt{E^2-P_y^2}},\quad b={K\over \sqrt{E^2-P_y^2}}
\ee
and the rescaled KK momentum 
\be
p={P_y \over \sqrt{E^2-P_y^2}} 
\ee
Setting $p=\sinh\gamma=s_\gamma$ so that $\sqrt{1+p^2} = \cosh\gamma= c_\gamma$, we have that the ubiquitous combinations  can be rewritten as
\begin{align}
&c_p \sqrt{1 + p^2} - s_p p = c_p c_\g - s_p s_\g = \cosh(\gamma - \delta_p) \equiv c_{\g - p} \\ 
&c_p p-\sqrt{1+p^2}s_p= c_ps_\gamma - s_pc_\gamma= \sinh(\gamma - \delta_p) \equiv s_{\gamma-p}
\end{align}
These are nothing but the relative `boost' parameters of the probe with respect to the background. The probe is, indeed, `charged', in that it carries KK momentum. This feature plays a crucial role in the analysis of the charge instability, in vague analogy with the charge super-radiance of Reissner-Nordstr{\"o}m black holes.
With all these identifications, the coefficients of \eqref{geopoly} read
\begin{align}
B=&-2r_-^2+r_+^2+{M\over2}(1+c_{2\d_1}+c_{2\d_5}+c_{2(\g-p)})-b^2\\\nn
C=&\Big\{b_\psi^2(a_1^2-a_2^2)+2Mb_\psi(a_1c_1 c_5 c_{\g-p}+a_2 s_1 s_5 s_{\g-p})+{a_1^2\over2}(r_+^2-r_-^2-2M)+\\\nn
&{+}{(a_1^2{-}a_2^2)^2\over 4}{+}(b_\psi\rightarrow b_\phi, a_1\leftrightarrow a_2)\Big\}{-}(r_+^2{-}r_-^2)b^2{+}{M(r_+^2-r_-^2)\over 2}(c_{2\d_1}{+}c_{2\d_5}{+}c_{2(\g-p)}){+}\\\nn
&{+}{M^2\over4}(3+c_{2\d_1}c_{2\d_5}+c_{2\d_1}c_{2(\g-p)}+c_{2\d_5}c_{2(\g-p)})\\\nn
D=&\Big\{{1\over2} b_\psi^2[(a_1^2-a_2^2)(r_+^2-r_-^2-a_1^2+a_2^2)+(a_1^2+a_2^2)M]+\\\nn
&+M b_\psi[a_1(M+r_+^2-r_-^2-a_1^2+a_2^2)c_1c_5c_{\g-p}-a_2(M-r_+^2+r_-^2+a_1^2-a_2^2)s_1s_5s_{\g-p}]+\\\nn
&{-}{a_1^2M^2\over 8}(c_{2\d_1}{+}c_{2\d_5}{+}c_{2(\g-p)}{+}c_{2\d_1}c_{2\d_5}c_{2(\g-p)}){-}{a_1a_2M^2\over8}s_{2\d_1}s_{2\d_5}s_{2(\g-p)}{+}(b_\psi\rightarrow b_\phi, a_1\leftrightarrow a_2)\Big\}{+}\\\nn
&{+}{M^2\over8}[(r_+^2-r_-^2+Mc_{2(\g-p)})(1+c_{2\d_1}c_{2\d_5})+(M+(r_+^2-r_-^2)c_{2(\g-p)})(c_{2\d_1}+c_{2\d_5})]
\end{align}

\subsection{Critical regime}

A particularly interesting class of geodesics are the critical ones, such that $\r=\r_c$. These are (possibly unstable) solutions of both $P_\r=0$ and $P'_\r=0$, or equivalently $Q_\r=0$ and $Q_\r'=0$. Critical geodesics form the so-called `light-ring' or `light-halo' for rotating geometries as JMaRT. Contrarily to the `event horizon', light-rings can be observed, possibly polluted by environmental effects, even for putative BHs such as M87* \cite{EventHorizonTelescope:2019dse} or SgrA* \cite{EventHorizonTelescope:2022wkp}.

Barring the denominator in $Q_\r$, that only vanishes at $\r^2=0$ where the geometry ends with a cap, the critical regime is determined by solving both the conditions for the vanishing of the numerator and of its derivative w.r.t. $\r$. 

In order to emphasize the dependence on the total impact parameter $b$, we rewrite the coefficients in \eqref{geopoly} as follows:
\be
B=\hat{B}-b^2,\quad C=\hat{C}-{a_f^2} b^2
\ee
$${a_f^2}=r_+^2-r_-^2\ge 0$$
while $D$ does not depend on $b$. The (un)stable light-rings satisfy
\be
{\cal R}_3(\r^2)={\cal R}_3'(\r^2)=0
\ee
which can be solved requiring that the cubic (in $\r^2$) polynomial ${\cal R}_3$ has a double root or zero discriminant, {\it viz.}
\be
\D_3(b_c^2)=18 B C D-4B^3D+B^2C^2-4C^3-27D^2=0
\ee
This is a quartic equation in $b^2$ that in principle allows to express $b^2$ in terms of $b_\phi, b_\psi$ and of the background parameters, subject to positivity constraints including $b\ge |b_\phi|+ |b_\psi|$, see below. 
The three roots are:
\be
\rho{^2}_{1,c}=-{B^3-4B C+9D\over B^2-3C }(b_c),\quad \rho{^2}_{2, c}=\rho{^2}_{3, c}=-{B C-9D\over2(B^2-3C)}(b_c)
\ee
The explicit form of the solutions for $\rho_c^2$ and $b^2_c$ is not very illuminating. The situation drastically simplifies for `shear-free' critical geodesics that can only exist for $\theta=0$, whereby $J_\phi=0$ and $J_\psi^2 = K^2 - (a_1^2-a_2^2)(E^2-P_y^2)$, or for $\theta=\pi/2$, whereby $J_\psi=0$ and $J_\phi^2 = K^2 - a_2^2(E^2-P_y^2)$. In the case of 2-charge circular fuzzballs, this kind of geodesics were studied in \cite{Bianchi:2022qph}. 

\subsection{Shear-free geodesics}
General (non critical) shear-free geodesics satisfy $\theta=\theta_0$ and $P_\theta=P'_\theta=0$. By imposing these conditions in \eqref{eqgeodA} one finds
\begin{align}
    &\hat{K}^2_{s{-}f}={J_\phi^2 \over \sin^2\theta_0}+ {J_\psi^2 \over \cos^2\theta_0}+ (a_1^2-a_2^2)\mathcal{E}^2 \cos^2\theta_0\\ &0= -{ J_\phi^2 \cos\theta_0 \over \sin^3 \theta_0}+ { J_\psi^2 \sin\theta_0 \over \cos^3 \theta_0}+ \sin \theta_0 \cos \theta_0 (a_1^2-a_2^2)\mathcal{E}^2 \sin\theta_0 \cos\theta_0
\end{align}
with $\hat{K}^2=K^2-a_2^2 \mathcal{E}^2$ and $\mathcal{E}^2=E^2-P_y^2$.\\Introducing the variables $\xi=\cos^2\theta_0, \beta^2={\hat{K}^2\over \mathcal{E}^2 (a_1^2-a_2^2)}, \beta_{\phi,\psi}^2={J_{\phi,\psi}^2 \over \mathcal{E}^2 (a_1^2-a_2^2)},$ one can rewrite the above equations in the form
\begin{align}
    &\beta^2={\beta_\phi^2 \over 1-\xi}+ {\beta^2_\psi \over \xi}+ \xi \\& \beta_\phi^2 {\xi \over 1-\xi}- \beta_\psi^2 {1-\xi \over \xi}+\xi(1-\xi)=0
\end{align}
Solving the system for $\beta_\phi$ and $\beta_\psi$ w.r.t $\xi$ and $\beta$ one finds
\be
\beta_\phi^2=(\beta^2 - 2\xi) (1-\xi)^2\quad , \quad \beta_\psi^2=(1+\beta^2 - 2\xi) \xi^2
\ee
For $\beta^2>2\xi$, $\theta_0$ must satisfy the constraint $\arccos {\hat{K} \over \mathcal{E} \sqrt{2 (a_1^2-a_2^2)}} < \theta_0< \pi - \arccos {\hat{K} \over \mathcal{E} \sqrt{2 (a_1^2-a_2^2)}}$.\\For $a_1\sim a_2$ one finds 
\be \beta_\phi^2\approx \beta^2 (1-\xi)^2\quad , \quad \beta_\psi^2\approx\beta^2 \xi^2
\ee
which implies that $\xi^2 \beta_\phi^2\approx\beta_\psi^2 (1-\xi)^2$ and therefore 
\be
\beta=|\beta_\phi| +|\beta_\psi| \quad \sim \quad \hat{K}=|J_\phi|+|J_\psi|
\ee
The wave counterpart of this condition is $\ell=|m_\phi|+|m_\psi|$, which is the minimal allowed value of $\ell$ for given $m_\phi$ and $m_\psi$, since in general $\ell=|m_\phi|+|m_\psi|+2h$ with $h$ a positive integer.

\subsection{Special geodesic at $\rho=0$}
\begin{figure}[h]
\centering
\includegraphics[width=8.5cm]{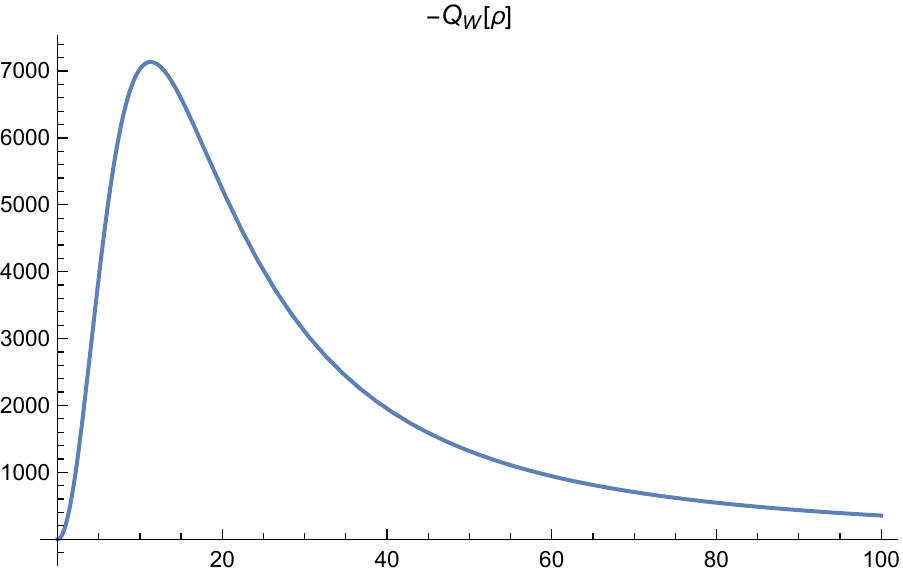}
\caption{Plot of $-Q_W(\rho)$ for $m=3$, $n=1$, $a_1=32$, $c_1=c_5=5$. For these parameters $b_c=1919.6$}\label{specialgeo}
\end{figure}

Setting $b_\phi=b_\psi=P_y=0$, $\rho=0$ turns out to be a critical geodesic for (see figure \ref{specialgeo}):
\begin{align}
 b_c^2=&\frac{1}{4 a_2^4 r_-^2 \left({\sigma}^4-1\right)}\Bigg\{\left(a_2^2+r_-^2\right) \left(a_2^2+r_-^2 {\sigma}^4\right) \left(2 a_2^2 r_-^2 \left({\sigma}^4-1\right)
   \left(c_p^2+s_p^2\right)+\right.\\\nn
   &\left.+\left(c_5^2+s_5^2\right) \left(\left(a_2^2+r_-^2\right) \left(a_2^2+r_-^2 {\sigma}^4\right) \left(c_p^2+s_p^2\right)+2 a_2^2
   r_-^2 \left({\sigma}^4-1\right)\right)+\right.\\\nn
   &\left.+\left(c_1^2+s_1^2\right) \left(\left(a_2^2+r_-^2\right) \left(a_2^2+r_-^2 {\sigma}^4\right)
   \left(c_p^2+c_5^2+s_p^2+s_5^2\right)+2 a_2^2 r_-^2 \left({\sigma}^4-1\right)\right)\right)+\\\nn
   &+4 a_2^2 r_-^6 \sigma^8+3 a_2^4 r_-^4 \left(\sigma^8+1\right)+4 a_2^6
   r_-^2 {\sigma}^4+a_2^8+r_-^8 {\sigma}^8\Bigg\}
\end{align}
where we recall that, according to $(\ref{parameters})$, ${\sigma}=\sqrt{s_1 s_5 s_p\over c_1 c_5 c_p}\leq1$.\\
This is the analogue of the critical geodesics along the circle $\rho=0$ of the circular D1-D5 fuzzball \cite{Bianchi:2022qph, Bianchi:2017sds}.

\subsection{General geodesics}
Taking $s_1=s_5=s$ for simplicity one has: 
\be
P_\r=\dot{\r}{(\r^2+r_+^2+M s^2+a_1^2\sin^2\q+a_2^2\cos^2\q)\over \r^2+r_+^2-r_-^2}
\ee
\be
P_\q=\dot{\q}(\r^2+r_+^2+Ms^2+a_1^2\sin^2\q+a_2^2\cos^2\q)
\ee
so that taking the ratio, that allows to cancel the intertwining factor $\r^2+r_+^2+Ms^2+a_1^2\sin^2\q+a_2^2\cos^2\q$, and integrating both sides between the initial and final points yields
\be\label{geoint}
\int_{\r_i}^{\r_f}{d\r \over P_\r(\r^2+r_+^2-r_-^2)}=\pm\int_{\theta_i}^{\theta_f}{d\q\over P_\q}
\ee
that could be expressed in terms of elliptic integrals.

For simplicity, setting $J_\psi=J_\phi=0$, we obtain:
\be
P_\q=\sqrt{K^2-\mathcal{E}^2(a_2^2\sin^2\q+a_1^2\cos^2\q)}
\ee
with $\mathcal{E}=\sqrt{E^2-P_y^2}$. Introducing the coordinate $y=\r^2$ and factorizing ${\cal R}_3(y)=\prod_{i=1}^3(y-y_i)$, \eqref{geoint} becomes:
\be
{\sqrt{K^2-a_1^2\mathcal{E}^2} dy\over 2\mathcal{E}\sqrt{\prod_{i=1}^3(y-y_i)}}=\pm{d\q\over \sqrt{1-{\mathcal{E}^2(a_2^2-a_1^2)\over K^2-a_1^2\mathcal{E}^2}\sin^2\q}}
\ee
Both these integrals can be written in terms of incomplete elliptic integrals of the first kind:
\begin{align}
\int_{\q_0}^{\q}{d\q'\over \sqrt{1-{\mathcal{E}^2(a_2^2-a_1^2)\over K^2-a_1^2\mathcal{E}^2}\sin^2\q'}}=\Big[F(\q;-{\mathcal{E}^2(a_1^2-a_2^2)\over K^2-a_1^2\mathcal{E}^2}-F(\q_0;-{\mathcal{E}^2(a_1^2-a_2^2)\over K^2-a_1^2\mathcal{E}^2}\Big]
\end{align}
and 
\begin{align}
\int_{y_0}^y{\sqrt{K^2{-}a_1^2\mathcal{E}^2} dy'\over 2\mathcal{E} \sqrt{\prod_{i=1}^3(y{-}y_i)}}
{=}{-}{\sqrt{K^2{-}a_1^2\mathcal{E}^2}\over 2\mathcal{E}\sqrt{y_{21}}}
\Big[F\left(\arcsin\sqrt{y_{21}\over y{-}y_1};{y_{13}\over y_{12}}\right){-}F\left(\arcsin\sqrt{y_{21}\over y_{01}};{y_{13}\over y_{12}}\right)\Big]
\end{align}
where $y_{ij}=y_i-y_j$. Inverting the radial or polar angular integrals in terms of Weierstrass $\wp$ function, allows in principle to fully solve the dynamics. In practice it is often convenient to rely on numerical methods.  

A simpler question  is to compute the polar angular deflection $\Delta\theta$ of a probe impinging from infinity reaching a simple turning point (with $P_\r=0$ but $P_\r'\neq 0$) and `bouncing' back to infinity. The radial integrals become `complete elliptic integrals' of the first kind.

\section{Wave equation}\label{sect4}
We are now ready to study the dynamics of a scalar wave perturbation in the background \eqref{eq:metric} and expose the sought for charge instability. We consider a scalar field with non-zero KK momentum $P_y$ that looks massive from a 5-dimensional perspective although it is massless from a 6-dimensional perspective.
The scalar wave equation then reads 
\begin{equation} \label{eq:waveeqn}
\Box_{(6)}\Phi = \frac{1}{\sqrt{- g}} \partial_\mu \bigl (\sqrt{-g} g^{\mu \nu} \partial_\nu \bigr ) \Phi = 0
\end{equation}
By defining $\chi=\cos\q$, the determinant of the metric is given by
\be
g=\det(g_{\mu \nu})=-r^2\sqrt{H_1 H_5} \chi^2 
\ee
The ansatz for the solution can be chosen to be
\be
\Phi=\exp\Big[i \bigl (-{{\omega}}\tilde{t}-{{P_y}}\tilde{y}+m_\phi\phi+m_\psi \psi \bigr )\Big]R(r)S(\chi)
\ee
This way, \eqref{eq:waveeqn} can be separated into a radial equation
\begin{align}
&{1\over r}{d\over dr}\Big[{(r^2-r_+^2)(r^2-r_-^2)\over r}{d\over dr}R(r)\Big]+\Big[{\omega^2-P_y^2\over R_y^2}(r^2+M s_1^2+Ms_5^2)+(\omega c_p+P_y s_p)^2{M}-\Lambda+\\\nn
&-(r_+^2-r_-^2){(P_y-n m_\psi+m m_\phi)^2\over r^2-r_+^2}+(r_+^2-r_-^2){(\omega \varrho+P_y \vartheta-n m_\phi+m m_\psi)^2\over r^2-r_-^2}\Big]R(r)=0
\end{align}
and an angular equation 
\begin{equation}
(1-\chi^2)S''(\chi)+{(1-3\chi^2)\over \chi}S'(\chi)+\Big[\hat{\Lambda}-{m_\phi^2\over 1-\chi^2}-{m_\psi^2\over \chi^2}+(\omega^2-P_y^2)(a_2^2-a_1^2)\c^2\Big]S(\chi)=0
\end{equation}
where
\be\label{hatLambda}
\varrho={c_1^2c_5^2c_p^2-s_1^2 s_5^2 s_p^2\over s_1 c_1 s_5 c_5},\quad \vartheta={c_1^2 c_5^2-s_1^2 s_5^2\over s_1 c_1 s_5 c_5}s_p c_p, \quad \hat{\Lambda}=\Lambda+(\omega^2-P_y^2)a_1^2
\ee
with $\omega=\tilde{\omega} c_p+\tilde{P_y} s_p$, $P_y=-(\tilde{P_y}c_p+\tilde{\omega}s_p)$ and $\Lambda$ the separation constant\footnote{The relation with the separation constant for geodesic motion is $K^2=\hat{\Lambda}+a_2^2(\omega^2-P_y^2)$.}. In this form the angular equation admits as eigenfunctions the oblate spheroidal harmonics in 5-dimensions \cite{Bianchi:2022qph, Berti:2005gp}.

Introducing the dimensionless coordinate
\be
x={r^2-r_+^2\over r_+^2-r_-^2}= {\rho^2 \over {a_f^2}}
\ee
the radial equation can be rewritten as
\be\label{errad}
{d\over dx}\Big[x(x+1){d\over dx}R(x)\Big]+{1\over 4}\Big[\kappa^2 x+1-\n^2+{\x^2\over x+1}-{\z^2\over x}\Big]R(x)=0
\ee
with
\begin{align}
&\kappa^2=(\omega^2-P_y^2)(r_+^2-r_-^2)\\\nn
&\x=\omega \varrho+P_y \vartheta-m_\phi n+m_\psi m\\\nn
&\z=P_y-m_\psi n+m_\phi m\\\nn
&\n^2=1+\Lambda-(\omega^2-P_y^2)(r_+^2+M s_1^2+Ms_5^2)-(\omega c_p+P_y s_p)^2{M}
\end{align}
Both radial and angular equations can be written in canonical form 
\be \label{eq:canonical}
\Psi''(z) + Q_W(z)\Psi(z) = 0
\ee
making use of 
\be
R(x)={\psi(x)\over \sqrt{x(1+x)}},\quad S(\c) = \frac{\Theta(\c)}{\sqrt{\c (1 - \c^2)}}
\ee
The radial $Q-$function becomes:
\be\label{eq:radialQ}
Q_{r,W}(x)={\kappa^2 x^3+(1+\kappa^2-\n^2)x^2+(1-\z^2-\n^2+\x^2)x+1-\z^2\over 4x^2(1+x)^2}
\ee
while the angular one written in terms of the coordinate $y = - \c^2$ is:
\begin{align} \label{eq:angularQ}
Q_{A,W}(y) =& \frac{1 - m_\psi^2}{4 y^2} + \frac{1 - m_\phi^2}{4 (1 + y)^2} + \frac{2 - m_\phi^2 - m_\psi^2 + \hat \Lambda - (\omega^2 - P_y^2)(a_1^2-a_2^2)}{4 (1 + y)} +\\\nn
&+\frac{-2 - \hat{\Lambda} + m_\phi^2 + m_\psi^2} {4 y}
\end{align}

Both equations can be written as RCHE (Reduced Confluent Heun Equation) with two regular singularities (in $0$ and $-1$) and one irregular singularity at infinity. They can be conveniently solved relying on their relation with quantum Seiberg-Witten (SW) curves for ${\cal N}=2$ SYM with $SU(2)$ gauge group and $N_f=(2,0)$ fundamental flavours. 

\section{Quantum Seiberg-Witten for QNMs}\label{sect5}
For completeness, we briefly describe the relation between BH and fuzzball Perturbation Theory and Quantum SW curves for ${\cal N}=2$ SYM with $SU(2)$ gauge group and $N_f=(N_L,N_R)$ fundamental flavours.
In the Coulomb branch, where the complex scalar in the vector multiplet gets a VEV $a=\langle\phi\rangle$, the low-energy dynamics is completely encoded in an analytic prepotential ${\cal F}(a)$, that only receives contributions at tree-level, one-loop and at the non-perturbative level from instantons. 

The scalar VEV $a$ and its magnetic dual $a_D=-{1\over 2\pi i} {\partial{\cal F}\over \partial a}$ can be viewed as periods of an elliptic curve  whose expression in the `commuting' variables $y$ and $x$ reads
\be
q P_L (x ) y + P_0(x) +P_R(x) y^{-1}=0
\ee
where $q=\Lambda^\beta$ with $\beta=2N_c-N_f=4-N_f$  is the instanton counting parameter and, in the case under consideration {\it i.e.} $N_f=(2,0)$, $\beta=2$ and 
\be
P_L(x) = (x-m_1) (x-m_2), \quad P_0(x)= x^2-u +q, \quad P_R(x)=1
\ee
with $u=a^2+ ... $ the gauge-invariant  Coulomb branch parameter.
In the `quantum' case, that corresponds to turning on a so-called $\Omega$-background \`a la Nekrasov-Shatasvili \cite{Nekrasov:2009rc}, $x$ and $y$ do not commute $[x,\log y]=\hbar$. Setting $x=\hbar y\partial_y$ leads to a second order differential equation of the same kind as the wave equation in JMaRT geometries that can be brought to canonical form, whereby 
the $Q$-functions \eqref{eq:radialQ} and \eqref{eq:angularQ} read 
\be \label{eq:SW}
Q_{2, 0} (y) = \frac{\hbar^2{-}(m_1{-}m_2)^2}{4\hbar^2 y^2} + \frac{\hbar^2{-} (m_1{+}m_2)^2}{4\hbar^2 (1 {+} y)^2} {+} \frac{\hbar^2{-}2 (m_1^2{+}m_2^2) {+} 4 u}{4\hbar^2 (1 + y)}{+}\frac{-\hbar^2{+}2 (m_1^2{+}m_2^2){+} 4 (q{-} u)}{4\hbar^2 y}
\ee

\subsection{Gauge/Gravity dictionaries}

This allows to establish precise gauge/gravity dictionaries, as follows.

For the angular equation one has
\be\label{dictA}
{m_{1,2}^A \over \hbar} = \frac{m_\phi \pm m_\psi}{2}, \quad {4u^A \over \hbar^2} =1+ \hat \Lambda - {(a_1^2 - a_2^2)(\omega^2-P_y^2)}, \quad {q^A\over \hbar^2} = - {1\over 4}{(a_1^2 - a_2^2) \ (\omega^2-P_y^2)}  
\ee
while for the radial equation one gets 
\be\label{raddict}
{m_{1, 2}^r \over \hbar} = {\x \pm \z \over 2} \quad, \quad {u^r \over \hbar^2} = {\k^2 + \n^2 \over 4} \quad, \quad {q^r\over \hbar^2} = {\k^2  \over 4} = {1\over 4}{(r_+^2 - r_-^2) \ (\omega^2-P_y^2)}  
\ee
They look similar since both of them correspond to $N_f=(2,0)$ qSW curves.

\subsection{$a$ and $a_D$ cycles }
We can also compute the quantum periods $(a, a_D)$ both for radial and angular equations. 
In the (2,0) flavour case, with  the above qSW curve,  
the SW differential reads
\be
\lambda_+(x)=-{x\over 2\pi i}{1\over W(x)}{dW(x)\over dx}
\ee
where $W(x)$ can be obtained recursively in powers of the instanton counting parameter $q$, using 
\be
q P_L\left(x-{\hbar\over 2}\right) P_R\left(x+{\hbar\over 2}\right) W(x)W(x-\hbar) + P_0(x) W(x) + 1 =0
\ee

The quantum period $a(u)$ can then be written as:
\begin{align}
a(u)&=\oint_\a \l_+=2\pi i\sum_{n=0}^\infty Res_{\sqrt{u}+n}\lambda_{+}(x)\\\nn
a(u)&=\sqrt{u}+{1-4m_{1}m_{2}-4u\over 4\sqrt{u}(4u-1)}q+\Bigg\{{2-3u\over64(u-1)u^{3/2}}+{3(m_{1}^2+m_{2}^2)\over16\sqrt{u}(1-5u+4u^2)}+\\\nn&+{m_{1} m_{2}(1-12u)\over4(1-4u)^2u^{3/2}}-{m_{1}^2m_{2}^2\Big[2+5u(12u-7)\Big]\over4(u-1)u^{3/2}(4u-1)^3}\Bigg\}q^2+\\\nn
&+\Bigg\{{5u-2-5u^2\over256(u-1)^2u^{5/2}}+{3(m_{1}^2+m_{2}^2)(1-15u+20u^2)\over64u^{3/2}(1-5u+4u^2)^2}+\\\nn&+{m_{1}m_{2}(m_{1}^2+m_{2}^2)(560u^3-1120u^2+497u-27)\over16(u-1)^2u^{3/2}(4u-9)(4u-1)^3}+\\\nn
&+{m_{1}m_{2}(54-861u+6397u^2-13100u^3+10480u^4-2880u^5)\over64(u-1)^2u^{5/2}(4u-9)(4u-1)^3}+\\\nn
&-{3m_{1}^2m_{2}^2(2-39u+371u^2-840u^3+560u^4)\over16(4u-1)^4(u-1)^2u^{5/2}}+\\\nn
&+{m_{1}^3m_{2}^3(18-413u+4705u^2-15260u^3+18480u^4-6720u^5)\over4(u-1)^2u^{5/2}(4u-9)(4u-1)^5}\Bigg\}q^3+{\cal O}(q^4)
\end{align}
Inverting this relation, one finds:
\begin{align}
\label{instseries}
&u=a^2+{4a^2+4m_{1}m_{2}-1\over2(4a^2-1)}q+\Bigg[{1\over 32(a^2-1)}-{3(m_{1}^2+m_{2}^2)\over 8-40 a^2+32a^4}+{(7+20a^2)m_{1}^2m_{2}^2\over2(a^2-1)(4a^2-1)^3}\Bigg]q^2+\\\nn
&+\Bigg[{5m_{1}m_{2}\over4(a^2-1)(4a^2-9)(4a^2-1)}{-}{m_{1}m_{2}(m_{1}^2{+}m_{2}^2)(17{+}28a^2)\over(4a^2{-}1)^3(a^2{-}1)(4a^2{-}9)}{+}{4m_{1}^3m_{2}^3(29{+}232a^2{+}144a^4)\over(4a^2{-}1)^5(a^2{-}1)(4a^2{-}9)}\Bigg]q^3 \\
&\quad +{\cal O}(q^4)
\end{align}
The correct boundary conditions (regularity and periodicity) for the angular dynamics require the quantization of the $a-$cycle:
\be
a_A=n_A+{1\over2}
\ee
Using the angular dictionary \eqref{dictA} with the definition \eqref{hatLambda}, we find:
\be
\hat{\Lambda}=\ell(\ell+2)-4q^A+4\sum_{k=1}^\infty u_k(q^A)^k
\ee
where we have identified $n_A=\ell/2$ and $u_k$ are the coefficients of the instanton expansion for $u_A$ \eqref{instseries}\cite{Bianchi:2022qph, Bianchi:2021mft}.

Let us focus on the case $\ell= 0$. Since $\ell\ge |m_\phi| + |m_\psi| $, then also $m_\phi = m_\psi = 0$. As a consequence, in \eqref{dictA}, $m^A_{1, 2} = 0$ and the cycle $a$ for the angular part reads\footnote{These last expressions are not general, they only hold for the angular part; we are omitting the apex $\theta$ just to ease the notation.}
\begin{align} 
\label{aAl0}
 a (u, q) & = \sqrt{u} -\frac{1}{4 \sqrt{u}} q + \frac{2-3 u}{64 (u-1)
   u^{3/2}} q^2 + \frac{5 u^2+5 u-2}{256 (u-1)^2 u^{5/2}} q^3 + \\\nn & + \frac{-175 u^4+875 u^3-952
   u^2+616 u-160}{16384 (u-4) (u-1)^3 u^{7/2}} q^4 + \\\nn & + \frac{
   -441 u^6+3822 u^5-11025 u^4+12828 u^3-10328 u^2+4672 u-896}{65536 (u-4)^2 (u-1)^4 u^{9/2}} q^5 + \mathcal{O} (q^6)
\end{align}
where we have expanded up to the fifth order in the angular coupling $q$. By inverting this relation, we end up with
\be
u = a^2 + \frac{1}{2} q + \frac{1}{32 (a^2 - 1)} q^2 + \frac{7 + 5 a^2}{8192(a^2 - 4)(a^2 - 1)^3} q^4 + \mathcal{O} (q^6)
\ee
For $\ell= 0$ we have $a = \frac{1}{2}$ and then 
\be 
u = \frac{1}{4} + \frac{q}{2} - \frac{q^2}{24} + \frac{11}{17280} q^4+\mathcal{O}(q^6)
\ee
We are ready to expose the presence of `charged' unstable QNMs with $\ell=0$ and $P_y\neq 0$.

\section{Charge instability: unstable modes}\label{sect6}
The origin of the instabilities of JMaRT geometries lies in the existence of an ergo-region without a horizon. These instabilities show themselves with the presence of scalar modes which are regular near the cap, behave as outgoing waves at infinity and grow in time. Furthermore we can distinguish between two kind of instabilities: angular \cite{Cardoso:2005gj} but also a charge-instability.

We would like to argue that a possible form of instability is the emission of KK charged (scalar) waves. In order to avoid any sort of `confusion' with other forms of ergo-region instability, in our analysis we will always consider $\ell=m_\phi=m_\psi=0$ and $P_y\neq 0$ and we will make use of different techniques to pin down the unstable QNMs. 

In this section, which represents the core of the present work, we compute charge instability modes by exploiting different techniques. Preliminarily we estimate the frequency of the mode by solving numerically the condition $a_D=0$ valid at one loop which actually coincides with matching the solution near the cap with the one at infinity as presented in \cite{Cardoso:2005gj} and summarized in appendix \ref{appA}. Then we will implement a numerical method of integration and matching and finally we will use the relevant quantization of SW cycles.
\subsection{1-loop SW}
Since the radial equation can be mapped to $(2,0)-$flavour quantum SW-curve with dictionary as in \eqref{raddict}, the condition $a_D=0$ with $a_D-$cycle valid at 1-loop is: 
\be\label{SW1loop}
    q^{-2a}+{\Gamma^2(1+2a)\Gamma({1\over2}-a+m_1)\Gamma({1\over2}-a+m_2)\over\Gamma^2(1-2a)\Gamma({1\over2}+a+m_1)\Gamma({1\over2}+a+m_2)}=0
\ee
In order to balance the term in $q$, the only way to solve the previous transcendental equation is to have one of the $\Gamma-$functions in the denominator blow up. We found that the correct argument to quantise is:
\be\label{quantcondSW}
{1\over2}+a+m_2=-N
\ee
with $N$ a non-negative integer. We proceed as follows: from \eqref{quantcondSW} together with the radial dictionary \eqref{raddict}, one could find an estimate for the real part of the mode and then use it as seed to solve numerically the full trascendental equation \eqref{SW1loop}.

\subsection{Numerical integration}
The numerical integration that we implemented exploits the ideas of the matched expansion in appendix \ref{appA}. In particular, starting from the cap $x=0$ $(\r=0, r^2=-|r_+|^2)$ with regular boundary condition given by \eqref{bcz0}, we numerically integrate the equation until a certain extraction point keeping the frequency as a matching parameter. Identically starting from the same extraction point we integrate until infinity where we impose the outgoing boundary condition \eqref{bcinf}. As a matching condition between these two solutions, we require their Wronskian to vanish. This condition can be read as an eigenvalue equation for the frequency of the unstable mode. In order to find the root of the Wronskian, we used the 1-loop SW estimate as a seed. 

\subsection{SW-quantization condition}
Using the connection formulae for RCHE \cite{Bonelli:2022ten} as in \cite{Bianchi:2022qph}, the relevant quantization condition compatible with outgoing waves at infinity is $a-a_D=n_r$, with $n_r$ the overtone number\footnote{Not to be confused with the integer $n$ characterizing the JMaRT solution.}. We will focus on the lowest overtone number $n_r=0$. Moreover we choose $a_1=19.1$,\quad $c_1=c_5=5$ as in \cite{Cardoso:2005gj}.
\subsection{Numerical results and discussion}
\begin{itemize}
\item $m=3$, $n=1$
\begin{table}[h!]
\begin{tabular}{|c|c|c|c|}
\hline
$P_y$ & 1-loop SW                      & Numeric                   & Seiberg-Witten          \\ \hline
$2$   & $3.63572 + {i}0.00294722 $ & $3.64386 + {i}0.000983264$ & $3.66415 + {i}0.0330258$ \\ \hline

$3$   & $4.52267 + {i}0.00531939$ & $4.53212 + {i}0.00234674 $ & $4.56034 + {i}0.0441793 $ \\ \hline

$4$   & $5.41148 + {i}0.0082198 $ & $5.42127 + {i}0.00473221 $ & $5.45765 + {i}0.0548001 $ \\ \hline

$5$   & $6.30238 +{i} 0.0114497 $  & $6.31181 + {i}0.00825624$  & $6.35608 +{i} 0.0646116$ \\ \hline

$6$   & $7.19551 + {i}0.0147874$ & $7.20429 +  {i}0.0126564$ & $7.25572 +  {i}0.0733879$ \\ \hline

$7$   & $8.09099 + {i}0.018012$ & $8.09899 + {i}0.0172887 $ & $8.15661 + {i}0.0809586$ \\ \hline

$8$   & $8.98887 + {i}0.0209189 $ & $8.99603 + {i}0.0213037 $ & $9.05884 + {i}0.0872107 $ \\ \hline

$9$   & $9.88918 + {i}0.0233283$ & $9.89557 + {i}0.0238925 $ & $9.96245 + {i}0.0920853$ \\ \hline

$10$  & $10.7919 + {i}0.0250881$  & $10.7977 + {i}0.0245029$   & $10.8674 +{i}0.0955708$ \\ \hline
\end{tabular}
\end{table}
\FloatBarrier
\item $m=11$, $n=8$
\begin{table}[h!]
\begin{tabular}{|c|c|c|c|}
\hline
$P_y$ & 1-loop SW                        & Numeric                    & Seiberg-Witten           \\ \hline
$2$   & $3.16901 + {i}0.0000208477$ & $3.1698 + {i}0.0000179592$ & $3.17162 + {i}0.00259428$ \\ \hline

$3$   & $3.89499 + {i}0.0000289205$ & $3.89604 + {i}0.0000242532 $ & $3.89814 + {i}0.00299483 $ \\ \hline

$4$   & $4.62165 + {i}0.0000315757$ & $4.62279 + {i}0.000028156 $ & $4.62498 + {i}0.00310297 $ \\ \hline

$5$   & $5.34913 + {i}0.0000249718 $ & $5.35005+{i}0.000027804$ & $5.35197 + {i}0.00285118$ \\ \hline

$6$   & $6.07766 + {i}6.80023*10^{-6}$ & $6.07805 + {i}0.0000161055$ & $6.07864 + {i}0.00176162 $ \\ \hline

\end{tabular}
\end{table}
\FloatBarrier
\item $m=17$, $n=11$
\begin{table}[h!]
\begin{tabular}{|c|c|c|c|}
\hline
$P_y$ & 1-loop SW                         & Numeric                     & Seiberg-Witten                                                     \\ \hline
$2$   & $3.41284 + {i}2.62404*10^{-6}$ & $3.41288 + {i}6.46364*10^{-6} $ & $3.41377 + {i}0.000963112$                                          \\ \hline

$3$   & $4.22001 + {i}4.08662*10^{-6}$ & $4.22007 + {i}9.26836*10^{-6} $ & $4.22118 + {i}0.00118401 $     \\ \hline

$4$   & $5.02732 + {i}5.44156*10^{-6} $ & $5.02757 + {i}0.0000118934$ & $5.0287 + {i}0.00134649 $     \\ \hline

$5$   & $5.83478 + {i}6.42295*10^{-6}$ & $5.83513 + {i}0.00001.40903$  & $5.83631 + {i}0.00144518$                                           \\ \hline
$6$   & $6.64241 + {i}6.74964*10^{-6}$ & $6.64276 + {i}0.0000155518$ & $6.64399 + {i}0.00147357 $     \\ \hline

$7$   & $7.45023 + {i}6.13288*10^{-6} $ & $7.45049 + {i}0.0000158717 $ & $7.45172 + {i}0.00142086 $     \\ \hline

$8$   & $8.25827 + {i}4.31724*10^{-6} $ & $8.25848 + {i}0.0000142972 $ & $8.25943 + {i}0.00125465 $     \\ \hline
    
$9$   & $9.06659 + {i}1.25859*10^{-6}$ & $9.06669 + {i}8.46263*10^{-6}$ & $9.067 + {i}0.000781647$\\ \hline
\end{tabular}
\end{table}
\FloatBarrier

\item $m=100$, $n=33$
\begin{table}[h!]
\begin{tabular}{|c|c|c|c|}
\hline
$P_y$ & 1-loop SW                         & Numeric                     & Seiberg-Witten                                                     \\ \hline
$2$   & $3.73952 + {i}1.1289*10^{-9} $ & $3.73862 +  {i}1.39197*10^{-7} $ & $3.73954 +  {i}0.0000210853 $     \\ \hline

$3$   & $4.65927 +  {i}1.92216*10^{-9} $ & $4.6578 +  {i}2.09551*10^{-7} $ & $4.6593 +  {i}0.0000273717 $     \\ \hline

$4$   & $5.57903 +  {i}2.86899*10^{-9} $ & $5.5769 +  {i}2.85962*10^{-7} $ & $5.57906 +  {i}0.0000332613 $     \\ \hline

$5$   & $6.49878 +  {i}3.93585*10^{-9} $ & $6.49625+{i}3.65687*10^{-7}$ & $6.49882 +  {i}0.0000387423 $     \\ \hline

$6$   & $7.41854 +  {i}5.08873*10^{-9} $ & $7.41589 +  {i}4.48409*10^{-7} $ & $7.41858 +  {i}0.0000438022 $     \\ \hline

$7$   & $8.3383 +  {i}6.29308*10^{-9} $ & $8.3351 +  {i}5.31804*10^{-7} $ & $8.33834 +  {i}0.0000484279 $     \\ \hline

$8$   & $9.25806 +  {i}7.51377*10^{-9} $ & $9.2548 +  {i}6.1437*10^{-7} $ & $9.25811 +  {i}0.0000526055 $     \\ \hline

$9$   & $10.1778 +  {i}8.71499*10^{-9} $ & $10.1746 +  {i}6.95964*10^{-7}$ & $10.1779 +  {i}0.0000563204 $     \\ \hline

$10$   & $11.0976 +  {i}9.86019*10^{-9} $ & $11.0941 +  {i}7.74257*10^{-7} $ & $11.0976 +  {i}0.000059557 $     \\ \hline

\end{tabular}
\end{table}
\FloatBarrier

\item $P_y=0$
\begin{table}[h!]
\begin{tabular}{|c|c|c|c|}
\hline
$(m,n)$        & 1-loop SW                         & Numeric                     & Seiberg-Witten            \\ \hline
$(3,1)$   & $1.86615 + {i}0.000284338$   & $2.15575 + {i}0.00013212$    & $1.87522 + {i}0.0104692$   \\ \hline
$(11,8)$  & $1.71893 + {i}3.18908*10^{-6}$ & $1.98708 + {i}6.63466*10^{-6}$ & $1.71989 + {i}0.00106042$  \\ \hline
$(17,11)$ & $1.79895 + {i}3.57665*10^{-7}$ & $2.07964 + {i}2.24714*10^{-6}$ & $1.79927 + {i}0.000365783$ \\ \hline
$(100,33)$ & $1.90001 + {i}1.3526*10^{-10} $ & $2.19557 + {i}4.57827*10^{-8} $ & $1.90002 + {i}7.36619*10^{-6} $ \\ \hline
\end{tabular}
\end{table}
\FloatBarrier
\end{itemize}
\FloatBarrier

Notice the clear appearance of unstable KK charged modes with ${\rm Im}\omega >0$ that signal the onset of charge instability in the JMaRT solution even for $\ell=0$. For classical waves charges are continuous, yet KK charges are quantized in terms of $R_y$. Emitting a quantum with minimal charge $P_y=1/R_y$ and (5-dimensional) mass $m=|P_y|$ the ADM mass $M_{ADM}$ and KK charge $Q_p$ of the JMaRT solution would reduce by the corresponding amounts $Q'_p=Q_p-P_y$ and $M'_{ADM}=M_{ADM}-|P_y|$. Due to spontaneous emission of charged quanta, the parameters of the solutions should rearrange so as to reach a nearby less unstable solution with different parameters. 

Setting 
\be
G_N^{(5)}= {\pi\over 4} = {G_N^{(10)}\over 2\pi R_y V_4}  = {8\pi^6 g_s^2 (\alpha')^4\over 2\pi R_y V_4}
\ee
the dressed charges  $Q_1$, $Q_5$ and $Q_p$ are related to the `quantized' charges $N_1$, $N_5$ and $N_p$ by 
\be
Q_1= {N_1 R_y \over g_s \alpha'}\quad , \quad Q_5= {N_5 R_y V_4\over 2 \pi^2 g_s \alpha' (2\pi\alpha')^2} \quad ,
\quad Q_p = {N_p\over R_y} = nm {Q_1 Q_5 \over R_y^2} 
\ee

The decrease of $N_p$ corresponds to a decrease of $nm$ since both $Q_1$ and $Q_5$ scale with $R_y$ at fixed $N_1$ and $N_5$. Small variations of $R_y$, $M$ and the related parameters $a_1$ and $a_2$ is necessary to reach a new smooth horizonless solution with $m'n'< mn$. 

\begin{figure}[h]
\centering
\includegraphics[width=8.5cm]{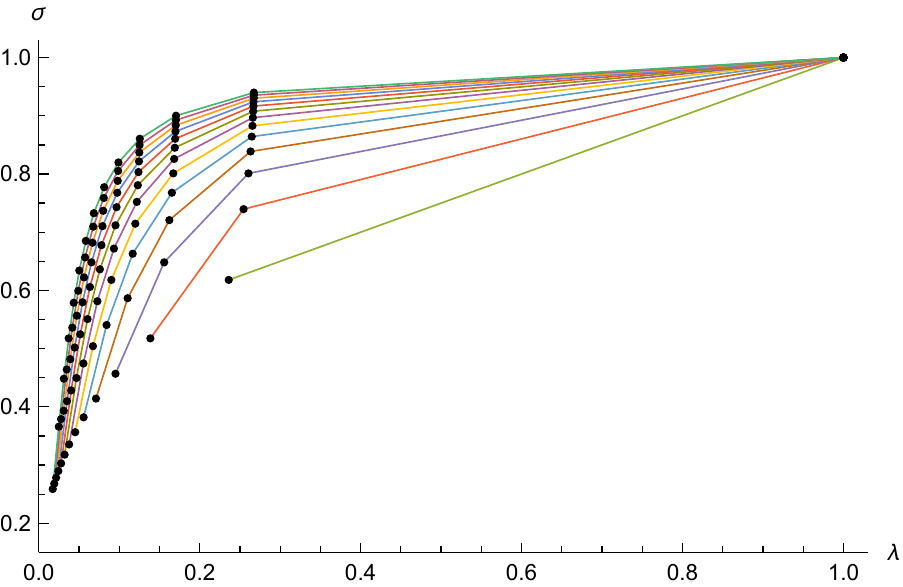}
\caption{Plot of the parameters $\lambda  = \sqrt{a_2/a_1}$ and 
$\sigma =\sqrt{\prod_i\tanh\delta_i}$ for different values of $m$ and $n$. Points connected by lines have the same value of $m$, ranging from $m=3$ (lowest line) to $m=15$ (highest line), with $n$ increasing from $n=1$ left-most to $n=m-1$ right-most.}
\label{plotjsigma}
\end{figure}

In Fig. \ref{plotjsigma} we display the parameters $\lambda  = \sqrt{a_2/a_1}$ and 
$\sigma =\sqrt{\prod_i\tanh\delta_i}$ for $m$ varying from $m=3$ to $m=15$ and $n$ increasing from $n=1$ to $n=m-1$.
 
In the simplest possible case ($m=3$, $n=1$ leftmost point on the lowest curve) the only JMaRT solution with a lower value of $mn$  is the BPS configuration ($m=2$, $n=1$ rightmost point) of the GMS family. The direct transition seems hard to achieve compatibly with angular momentum conservation ($\ell=0$), since it requires a significant variation of $\lambda=\sqrt{a_2/a_1}$ and $\sigma = \sqrt{\prod_i\tanh\delta_i}$. This transition may require the emission of charged (or neutral) quanta with $\ell\neq 0$. However, for larger values of $m$ and $n$ many points in the diagram lie close to one another. The variations are small and the transitions can be more likely to take place by emission of charged quanta with $\ell=0$.

Similar discharge mechanisms should be possible for the other charges $Q_1$ and $Q_5$ that couple to wrapped D-strings and D5-branes. Even at the linearized level, setting up the stage for the study of D-brane perturbations looks rather challenging. On the other hand computing the rate of emission of KK charged (BPS) quanta looks feasible using the by-now available connection formulae for the RCHE \cite{Bonelli:2022ten}.

\section{Conclusions and outlook} \label{sect7}

We argued that JMaRT solution admit a class of unstable KK charged QNMs with $\ell=m_\phi=m_\psi=0$, and $P_y\neq 0$. This suggests that charge instability is one of the mechanism of decay to less unstable or BPS solutions, that are known to be stable at the linear level but may suffer from non-linear instabilities due to the presence of an evanescent ergo-region \cite{Eperon:2016cdd}. 

Similar decay processes involving emission of wrapped D1's or D5's should also contribute but are much harder to analyze in terms of wave equations or S-matrix even at the linearized level. U-duality can be invoked to argue for a substantial equivalence among the various decay channels. 

More importantly, the very instability of topological solitons such as JMaRT may cast some doubts on the stability of other solutions of this kind but with lower angular momentum \cite{Heidmann:2022ehn}. The hope is that over-rotation is the culprit and when the solution has low or zero angular momentum, and thus no ergo-region, it may be stable. However charge and spin are two faces of the same coin if one allows for `dimensional' oxidation and reduction of the solution \cite{Aalsma:2022knj} and the story might be more involved. 

We hope to report on these issues in the near future \cite{TopStarStab}. 

\section*{Acknowledgements}
We acknowledge fruitful scientific exchange with I.~Bena, G. Bonelli, G. Bossard, V.~Cardoso, D. Consoli,
G.~Dibitetto, F. Fucito, A. Grillo, D.~Mayerson, F. Morales, P.~Pani, R. Savelli and A. Tanzini.
M.~B. and G.~D.~R. would like to thank GGI Arcetri (FI) for the kind hospitality during completion of this work. We thank the MIUR PRIN contract
2020KR4KN2 ``String Theory as a bridge between Gauge Theories and Quantum Gravity'' and the INFN project ST\&FI ``String Theory and Fundamental Interactions'' for partial support.

\appendix
\section{Angular superradiance}\label{appA}
Before analysing charged unstable modes, we calibrated our algorithms by reproducing some results already present in literature \cite{Cardoso:2005gj}. 
In this appendix, we briefly summarize the matching procedure adopted out in \cite{Cardoso:2005gj}. Near the cap we can neglect the term $\kappa^2 x \ll |1-\n^2|$ in \eqref{errad}, so that:
\be
x(1+x)R''(x)+(1+2x)R'(x)+{1\over 4}\Big[1-\n^2+{\x^2\over x+1}-{\z^2\over x}\Big]R(x)=0
\ee
which is actually a Gaussian hypergeometric equation. Given its solutions
\be\label{bcz0}
R(x)=Ax^{|\z|/2}(1+x)^{\x/2}{}_2F_1(a,b,c,-x)+Bx^{-|\z|/2}(1+x)^{\x/2}{}_2F_1(a-c+1,b-c+1,2-c,-x)
\ee
$$a={1\over 2}(1+|\z|+\z+\n),\quad b={1\over 2}(1+|\z|+\x-\n),\quad c=1+|\z|$$
we have to impose their regularity in $x=0$, since the geometry has there a smooth cap. The singular term in $x=0$ is eliminated by taking $B=0$.

At large $x$, by making use of the hypergeometric connection formulae, the near-cap solution will appear as:
\begin{align}\label{math2}
R(x)=&A\Gamma(1+|\z|)\Bigg[{\Gamma(-\n)\over \Gamma[{1\over2}(1+|\z|+\x-\n)]\Gamma[{1\over2}(1+|\z|-\x-\n)]}x^{-{\n+1\over2}}+\\\nn
&+{\Gamma(\nu)\over \Gamma[{1\over 2}(1+|\z|+\x+\nu)]\Gamma[{1\over 2}(1+\z-\x+\nu)]}x^{{\n-1\over2}}\Bigg]
\end{align}
For $x \gg 1$ the radial equation can be approximated as follows:
\be
R''(x)+{2\over x}R'(x)+\Big[{\kappa^2\over 4x}-{\nu^2-1\over 4x^2}\Big]R(x)=0
\ee
whose solution can be written in terms of the Bessel function of the first kind:
\be\label{bcinf}
R(x)=x^{-1/2}\Big[C J_\nu(\kappa k\sqrt{x})+DJ_{-\nu}(\kappa \sqrt{x})\Big]
\ee
Since we are interested in outgoing wave condition at infinity, the previous solution can be expanded for $\kappa\sqrt{x} \gg 1$ as
\be\label{infbehaviour}
R(x)\sim {x^{-3/4}\over \sqrt{2\pi \kappa}}\Big[e^{i\kappa \sqrt{x}-i{\pi\over4}}(C e^{-i{\pi \nu\over 2}}+De^{i{\pi \nu\over2}})+e^{-i\kappa\sqrt{x}+i{\pi\over4}}(Ce^{i{\pi\nu\over 2}}+De^{-i{\pi\nu\over2}})\Big]
\ee
Since in \cite{Cardoso:2005gj} the real part of omega is negative, the correct choice of the coefficients $C$ and $D$ that are compatible with outgoing wave condition at infinity is:
\be
C=-De^{i\pi\nu}
\ee
Plugging the previous condition in \eqref{infbehaviour} and expanding for small $\kappa\sqrt{x}$, we obtain
\be\label{math1}
R(x)\sim D\Bigg[{(2/ \kappa)^{-\nu}\over\Gamma(1+\nu)}x^{\nu-1\over2}-e^{i\pi\nu}{(2/\kappa)^\nu\over\Gamma(1-\nu)}x^{-{\nu+1\over2}}\Bigg]
\ee
By matching \eqref{math2} with \eqref{math1}, we obtain:
\be\label{mathching}
e^{i\pi\nu}(\kappa/2)^{2\nu}={\Gamma(1+\nu)^2\Gamma[{1\over2}(1-\nu+|\z|+\x)]\Gamma[{1\over2}(1-\nu+|\z|-\x)]\over \Gamma(1-\nu)^2\Gamma[{1\over2}(1+\nu+|\z|+\x)]\Gamma[{1\over2}(1+\nu+|\z|-\x)]}
\ee
Since the left hand side of \eqref{mathching} is suppressed by the presence of $\kappa$, at first instance, we are forced to take one of the $\Gamma-$functions in the denominator at right-hand side of \eqref{mathching} to be large. The correct choice is:
\be\label{realomegaWKB}
\nu+|\z|-\x=-(2N+1)
\ee
where $N$ is a non negative integer. In our numerical computation, since we notice that the absolute value of the real part is in general much bigger than the absolute value of the imaginary part, we used the solution of \eqref{realomegaWKB} as a seed for the resolution of \eqref{mathching}. A different procedure is perfomed in \cite{Cardoso:2005gj} where they provide a perturbative estimation of the imaginary part. As for charge instability modes, we used the result of this matching procedure as a seed for our numerical algorithm and for SW-quantization condition.

Fixing the parameters as in \cite{Cardoso:2005gj}, {\it viz.}
\be
m=5,\quad n=1,\quad c_1=1.1,\quad c_5=1.52,\quad a_1=262.7,\quad P_y=0,\quad \ell =m_\psi,\quad m_\phi=0
\ee
we obtain the angular instability modes \eqref{tabangular}. These are perfectly consistent with those computed in \cite{Cardoso:2005gj} which we show in \ref{tabangular2} for an easier comparison. The small differences between the two tables \eqref{tabangular} and \eqref{tabangular2} come from different choices of the positions of the numerical cap and radial infinity.  
\begin{table}[h]\small
\begin{tabular}{|c|c|c|c|}
\hline
$m_\psi$ & Matching                          & Numeric                                        & Seiberg-Witten                \\ \hline
$1$      & $-0.184725 + {i}3.98659*10^{-8} $ & $-0.184428 + {i}2.18488*10^{-8} $                  & $-0.184725 + {i}3.9815*10^{-8} $ \\ \hline
$2$      & $-0.752333 + {i}5.23579*10^{-8} $ & $-0.752694 + {i}3.51363*10^{-8} $                  & $-0.752333 + {i} 3.47118*10^{-8}$  \\ \hline
$3$     & $-1.32971 + {i}6.41976*10^{-9} $ & $-1.33053 + {i}5.99006*10^{-9} $                  & $-1.32971 + {i}6.34331*10^{-9}$ \\ \hline
$4$      & $-1.91174 + {i}9.41491*10^{-10} $  & $-1.9131 + {i}1.07648*10^{-9} $                  & $-1.91174 + {i}9.23543*10^{-10}$  \\ \hline
$5$      & $-2.49642 + {i}1.28253*10^{-10} $ & $-2.49838 + {i}1.27252*10^{-10} $ &               $-2.49642 + {i}1.24827*10^{-10}$ \\ \hline
$6$      & $-3.08276 + {i}1.69065*10^{-11} $  & $-3.08535 + {i}1.785*10^{-11} $                  & $-3.08276 + 1.63202*10^{-11}$  \\ \hline
$7$      & $-3.67023 + {i}2.18982*10^{-12} $  & $-3.67348 + {i}2.47565*10^{-12} $                   & $-3.67023 + {i}2.09597*10^{-12}$  \\ \hline
$8$      & $-4.25849 + {i}2.80624*10^{-13} $ & $-4.26242 + {i}2.96179*10^{-13} $                  & $-4.25849 + {i}2.66263*10^{-13}$ \\ \hline
$9$      & $-4.84734 + {i}3.57071*10^{-14} $ & $-4.84734 + {i}3.78729*10^{-14} $                  & $-4.84734 + {i}3.35795*10^{-14}$ \\ \hline
$10$     & $-5.43663 + {i}4.52044*10^{-15} $ & $-5.44197 + {i}5.31335*10^{-15} $                  & $-5.43663 +{i} 4.52044*10^{-15} $ \\ \hline
\end{tabular}
\caption{}\label{tabangular}
\end{table}
\begin{table}[]\small
\centering
\begin{tabular}{|c|c|c|}
\hline
$m_\psi$ & Matching                          & Numeric                 \\ \hline
$1$      & $-0.184+{i}3.83*10^{-8} $ & $-0.184+{i}3.83*10^{-8} $ \\ \hline
$2$      & $-0.744+{i}251*10^{-8} $ & $-0.744+{i}2.64*10^{-6} $   \\ \hline
$3$     & $-1.312+{i}3.37*10^{-9} $ & $-1.312+{i}3.53*10^{-9} $ \\ \hline
$4$      & $-1.883+{i}3.69*10^{-10} $  & $-1.882+{i}3.63*10^{-10} $   \\ \hline
$5$      & $-2.546+{i}3.55*10^{-11} $ & $-2.454+{i}3.39*10^{-11} $  \\ \hline
$6$      & $-3.030+{i}3.22*10^{-12} $  & $-3.028+{i}3.02*10^{-12} $    \\ \hline
$7$      & $-3.605+{i}2.77*10^{-13} $  & $-3.602+{i}2.63*10^{-13} $    \\ \hline
$8$      & $-4.180+{i}2.47*10^{-14} $ & $-4.176+{i}2.24*10^{-14} $   \\ \hline
$9$      & $-4.755+{i}2.05*10^{-15} $ & $-4.751+{i}1.89*10^{-15} $  \\ \hline
$10$     & $-5.331+{i}1.76*10^{-16} $ & $-5.326+{i}1.58*10^{-16} $   \\ \hline
\end{tabular}
\caption{}\label{tabangular2}
\end{table}
\begin{figure}[h]
\centering
\includegraphics[width=8.5cm]{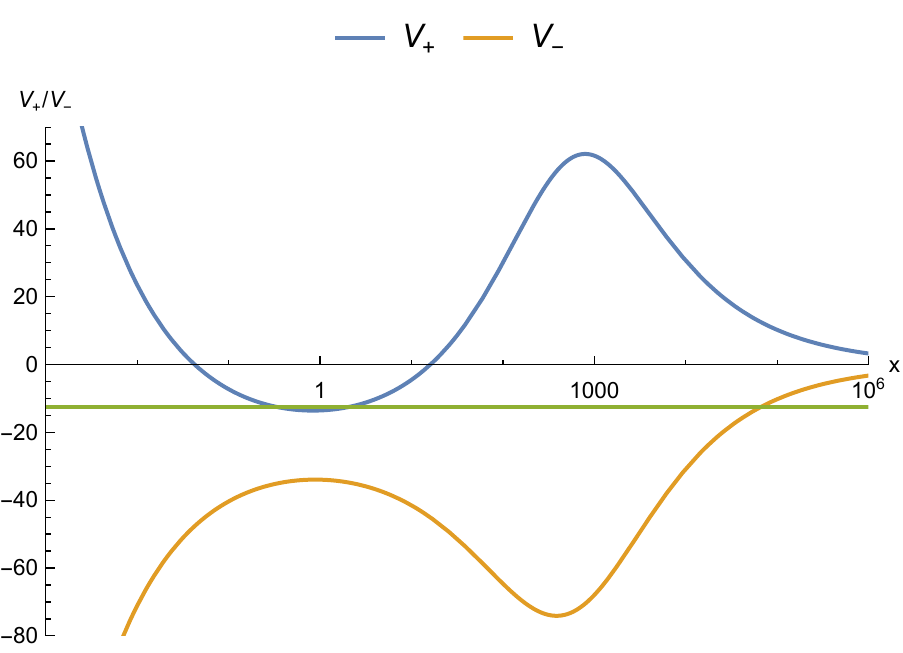}
\caption{Plot of the effective potentials read by $\psi''(z)+f(\rho)(\omega-V_+)(\omega-V_-)=0$ with parameters $m=5$, $n=1$, $a_1=32$, $c_1=c_5=5$,$P_y=0$, $m_\phi=0$ and $l=m_\psi=5$. With this parameters $\omega_R=-12.4791$ which is represented by green line. As we can see, the mode is nearly-stable for $V_+$, but the tunneling to infinity is allowed by $V_-$.}\label{potpimezzieqplane}
\end{figure}
\FloatBarrier

\bibliography{bibJMaRT}
\end{document}